\documentclass[aps,prb,twocolumn,groupedaddress,showpacs,showkeys]{revtex4}
\usepackage{graphicx}

\input{tcilatex}
\begin{document}

\title{Is planetary chaos related to evolutionary (phenotypic) rates?}
\author{J. C. Flores }
\affiliation{Instituto de Alta Investigaci\'{o}n IAI, Universidad de Tarapac\'{a},
Casilla 7-D, Arica, Chile}

\begin{abstract}
After Laskar\cite{laskar1,laskar2}, the Lyapunov time $\tau $ in the solar
system is about five millions years ($\tau =5.000.000$ [years]). On the
other hand, after Kimura\cite{kimura}, the evolutionary (phenotypic) rate $%
\nu $, for hominids, is $\nu =1/5.000.000$ [1/years]. Why are these two
quantities so closely related ($\nu \tau =1$)? In this work, following a
proposition by Finlayson\cite{fin} and Hutchings\cite{hut} \textit{et al}, I
found an inequality, which relates Lyapunov time and evolution rate. This
inequality fits well with some known cases in biological evolution.
\end{abstract}

\pacs{05.45.Gg, 87.18.Tt, 87.23.Kg, 05.45.-a}
\keywords{chaos, biological systems, evolution}
\maketitle

\section{Introduction}

Chaos plays a major role in the modern conception of physics\cite{scott},
particularly related to the instability of the planetary system. On the
other hand, phenotypic evolution plays an important role in biological
sciences. Surprisingly this two different aspect of research are closely
related since the rate of evolution (for some superior species) \ is closely
related to the Lyapunov exponent.

\section{Inequality for evolution rate and Lyapunov time}

Following Finlayson\cite{fin} (page 42), and referring to human evolution,
one way to reduce the effect of environmental fluctuations is to prolong the
response time face to these changes (based on an original idea by Hutchings%
\cite{hut} \textit{et al}). In fact, a slow evolutionary response may be
able to keep a stable population. This is very natural since face to
fluctuations, stability becomes related to long time delay.

After these ideas, consider the proposed chaotic instability for the solar
system (see reference 6 for a general discussion). That is, small
perturbations produce unpredictable changes after a time $\tau $ called the
Lyapunov time. In the solar system this time was evaluated numerically\cite%
{laskar1,laskar2,scott} and corresponds to $\tau =5.000.000$ [years]. It
means that the solar system possesses an intrinsic noise source due to chaos
and characterized by that time. Following Hutchings \textit{et al} and
Finlayson ideas, one expects that the evolution-time ($1/\nu $) is bigger
than the fluctuations-time related to chaos instabilities. That is,

\begin{equation}
\nu \tau \leq 1,  \label{desi}
\end{equation}%
and in the particular case of the planetary system

\begin{equation}
\nu \leq 1/5\text{ [darwins].}  \label{darwin}
\end{equation}%
Where one darwin corresponds to 1/1.000.000 [1/years]. Some values of the
evolution rates are $\nu =0.04$ [darwins] for horses, and $\nu =0.026$
[darwins] for a few species of dinosaurs in the Mesozoic\cite{kimura}. Note
that (\ref{darwin}) it is not contradictory with a low evolutionary rate
(crocodile and other with small rates). Nevertheless, it gives a superior
bound for evolution rates guided by noise in the solar systems. This is the
case for hominids (after the Kimura's rate\cite{kimura}, $\nu =1/5$
[darwins]) where the equality is verified in (\ref{darwin}).

\section{Conclusions}

I have proposed an inequality, based on the ideas of Hutchings et al and
Finlayson, which relates the evolution rate and environmental (noise)
instability-time. Particularly, for some species (including hominids), this
inequality is well verified when the planetary instability due to chaos is
considered. Naturally, for species with high evolutionary rate (for instance
virus) the origin of fluctuations is a more local (terrestrial) source.

A final comment, inequality (\ref{desi}) ensures stability face to
environmental fluctuations (noise) nevertheless face to abrupt and marked
changes, in the sense of catastrophe theory, a slow response does not
guarantee stability.

\begin{acknowledgments}
This work was supported by project FONDECYT 1070597, and UTA-Mayor Project
(2005-2006). I acknowledge K. J. Chandia who showed me reference 4.
\end{acknowledgments}

\end{document}